\documentclass[runningheads]{llncs}
\usepackage{graphicx}
\usepackage{subfigure}
\usepackage{algorithm} 
\usepackage{algpseudocode}
\usepackage{array}
\usepackage{epsfig}
\usepackage{amsmath}
\usepackage{hyperref}

\hypersetup{
colorlinks=true,
linkcolor=blue,
filecolor=magenta, 
urlcolor=blue,
citecolor=blue,
pdftitle={Sharelatex Example},
bookmarks=true,
pdfpagemode=FullScreen,
}
\begin{document}

\title{Automatic Frame Selection Using MLP \\Neural Network in Ultrasound Elastography
\thanks{This research was funded by Richard and Edith Strauss Foundation.}
}

\titlerunning{MLP NN Frame Selection in Elastography}
%
\author{Abdelrahman Zayed \and
Hassan Rivaz}

\authorrunning{A. Zayed and H. Rivaz}
%
\institute{Department of Electrical and Computer Engineering,\\
PERFORM Centre, \\
Concordia University, Montreal, QC H3G 1M8, Canada \\
\email{a\_zayed@encs.concordia.ca, hrivaz@ece.concordia.ca}\\ 
}
\maketitle 

\begin{abstract}
Ultrasound elastography estimates the mechanical properties of the tissue from two Radio-Frequency (RF) frames collected before and after tissue deformation due to an external or internal force. This work focuses on strain imaging in quasi-static elastography, where the tissue undergoes slow deformations and strain images are estimated as a surrogate for elasticity modulus. The quality of the strain image depends heavily on the underlying deformation, and even the best strain estimation algorithms cannot estimate a good strain image if the underlying deformation is not suitable. Herein, we introduce a new method for tracking the RF frames and selecting automatically the best possible pair. We achieve this by decomposing the axial displacement image into a linear combination of principal components (which are calculated offline) multiplied by their corresponding weights. We then use the calculated weights as the input feature vector to a multi-layer perceptron (MLP) classifier. The output is a binary decision, either 1 which refers to good frames, or 0 which refers to bad frames. Our MLP model is trained on \textit{in-vivo} dataset and tested on different datasets of both \textit{in-vivo} and phantom data. Results show that by using our technique, we would be able to achieve higher quality strain images compared to the traditional methods of picking up pairs that are 1, 2 or 3 frames apart. The training phase of our algorithm is computationally expensive and takes few hours, but it is only done once. The testing phase chooses the optimal pair of frames in only $1.9$ ms. 

\end{abstract}
\section{Introduction}
Ultrasound elastography is a branch of tissue characterization that aims to determine the stiffness of the tissue. Elastography has a significant potential in improving both detection and guiding surgical treatment of cancer tumors since tumors have higher stiffness values compared to the surrounding tissue \cite{ref_article1}. Elastography can be broadly divided into dynamic and quasi-static elastography \cite{ref_article2}, where the former deals with faster deformations in the tissue such that dynamics of motion should be considered. In this paper, we focus on quasi-static elastography, and in particular, quasi-static strain imaging where the final goal is to estimate strain images. In quasi-static elastography, tissue deformations are slow and therefore motion dynamics can be ignored.

In spite of the wide range of applications that quasi-static elastography has, it is highly user-dependent, which has hindered its widespread use. 
A pure axial compression yields higher quality strain images compared to a compression that has both in-plane and out-of-plane displacements. Therefore, the user needs to be highly skilled in axially deforming the tissue. Even for highly skilled users, some organs are hard to reach and the probe needs to be held in angles and directions that make imaging yet more challenging. Therefore, it has become crucial to develop a method for selecting the frames that result in strain images of high quality.

In order to make the strain image quality independent of the experience the user has in applying purely axial compression, Hiltawsky el al. \cite{ref_article3} developed a freehand applicator that can apply purely axial force regardless of the user's experience. The transducer could be put on a fixed surface moving vertically in the range of 1 to 2 mm.

Jiang at al. \cite{ref_article4} worked on frame selection by defining a quality metric for performance assessment and maximizing it. This metric depends on the normalized cross correlation (NCC) between Radio-Frequency (RF) frames and the NCC between their corresponding strain images. 

Another approach by Foroughi et al. \cite{ref_article5} used an external tracker that gives complete information about the location of the RF frame at the time of being produced, where frames collected from the same plane are selected. Among the selected frames, they only chose some of them according to a defined cost function that maximized axial compression. 

Although the previously mentioned approaches showed an improvement over the traditional way of picking up RF frames while maintaining a fixed gap between them, they also have some drawbacks, such as the need for an external mechanical applicator \cite{ref_article3} or an external tracking device \cite{ref_article5}. Other approaches such as \cite{ref_article4} need to calculate the strain before determining whether the pair of frames is good or not, so we can't use it in real-time applications, especially when we have a search range for finding good frames.

Herein, we introduce a novel real-time method for determining good RF frames used to obtain high-quality strain images, without the need of any external hardware. In the training phase, we calculate a set of principal components for quasi-static elastography. In the test phase, we develop a fast technique to find any compression as a weighted sum of those principal components. We then develop a Multi-Layer Perceptron (MLP) Neural Network to classify each pair of RF data as suitable or unsuitable for elastography.

\section{Methodology}
\begin{figure}[H]
\begin{center}
\subfigure{\includegraphics[height=4.6 cm,width=4.5 cm]{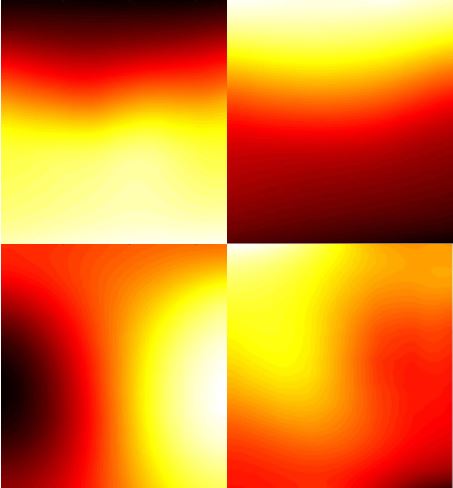}}
\end{center}%
\caption{Principal components of in-plane axial displacement learned from both \textit{in-vivo} and phantom experiments. Top row represent desirable axial deformation principal components.} \label{fig0}
\end{figure}

Let two RF frames $I_1$ and $I_2$ be collected before and after some deformation in the tissue. Our goal is to determine whether or not they are suitable for strain estimation. However, developing a classifier that takes the RF frames as an input and outputs a binary decision is not practical, as the number of samples in each RF frame is approximately one million, and therefore, a large network with a powerful GPU is required~\cite{ref_article6,ref_article7}. To solve the problem, we calculate $N$ principal components that describe the axial displacement as the tissue deforms. These principal components are represented by $\textbf{b}_1$ to $\textbf{b}_{N}$. Fig. \ref{fig0} shows some of these principal components learned from real experiments. We then calculate a coarse estimation of the axial displacement that occurred to the pixels between the two frames using Dynamic Programming (DP) \cite{ref_article8}, where we only get an integer value of the axial displacement. Due to the computational complexity of DP, we don't run it on the whole RF image, it is only run on a very small number of RF lines to get their displacement. After that we decompose the displacement into a linear weighted combination of the principal components that we computed offline. The resulting weight vector corresponds in a one-to-one relationship with the displacement image, but it has a lower dimensionality, which means that we can use it as the input to a multi-layer perceptron (MLP) classifier.

\subsection{Feature extraction}

Let the dimensions of each of the RF frames $I_1$ and $I_2$ be $m\times l$, where $m$ refers to the number of samples in an RF line and $l$ is the number of RF lines. We start by choosing $p$ equidistant RF lines (where $p < < l$), then we run DP to get their integer displacement values, resulting in $K$ estimates (where $K = m \times p$). We then form a $K$-dimensional vector \textbf{c} that has the displacement estimates of only a few sparse points out of the total $m \times l$ that we have in the RF image.
In the next step, we form the matrix \textbf{A} such that
\begin{equation}
\textbf{A}=
\begin{bmatrix}
\textbf{ b}_1(q_1) &\textbf{ b}_2(q_1) & \textbf{ b}_3(q_1) & \dots &\textbf{ b}_N(q_1) \\
\textbf{ b}_1(q_2) & \textbf{ b}_2(q_2) & \textbf{ b}_3(q_2) & \dots & \textbf{ b}_N(q_2) \\
\hdotsfor{5} \\
\textbf{ b}_1(q_K) & \textbf{ b}_2(q_K) & \textbf{ b}_3(q_K) & \dots & \textbf{ b}_N(q_K)
\end{bmatrix}
\end{equation}
where $q_1$ to $q_K$ refer to the 2D coordinates of our $K$ sparse points chosen along the $p$ RF lines. We then solve the optimization equation below:
\begin{equation}
\label{lsq}
\hat{\textbf{w}}= \textrm{arg}\min_\textbf{w} ||\textbf{Aw--c}|| 
\end{equation}
This means that the linear combination of the $N$ principal components multiplied by the weight vector $\textbf{w}=(w_1,...,w_N)^T$ would result in the displacement image with the minimum sum-of-squared error. Algorithm 1 summarizes the procedure for feature extraction.
\begin{algorithm}[H]
\caption{}
\begin{algorithmic}[1]
\Procedure {~}{}
\State \text{Choose $p$ equidistant RF lines}
\State \text{Run DP to get the integer axial displacement of the $p$ RF lines}
\State \text{Solve Eq.~\ref{lsq} to get the vector $\textbf{w}$}
\State \text{Pass the vector $\textbf{w}$ as input to the MLP classifier}
\EndProcedure
\end{algorithmic}
\label{algorithm:pca_glue}
\end{algorithm}

\subsection{Training the MLP Classifier}

\begin{figure}
\centering
\includegraphics[width=\textwidth]{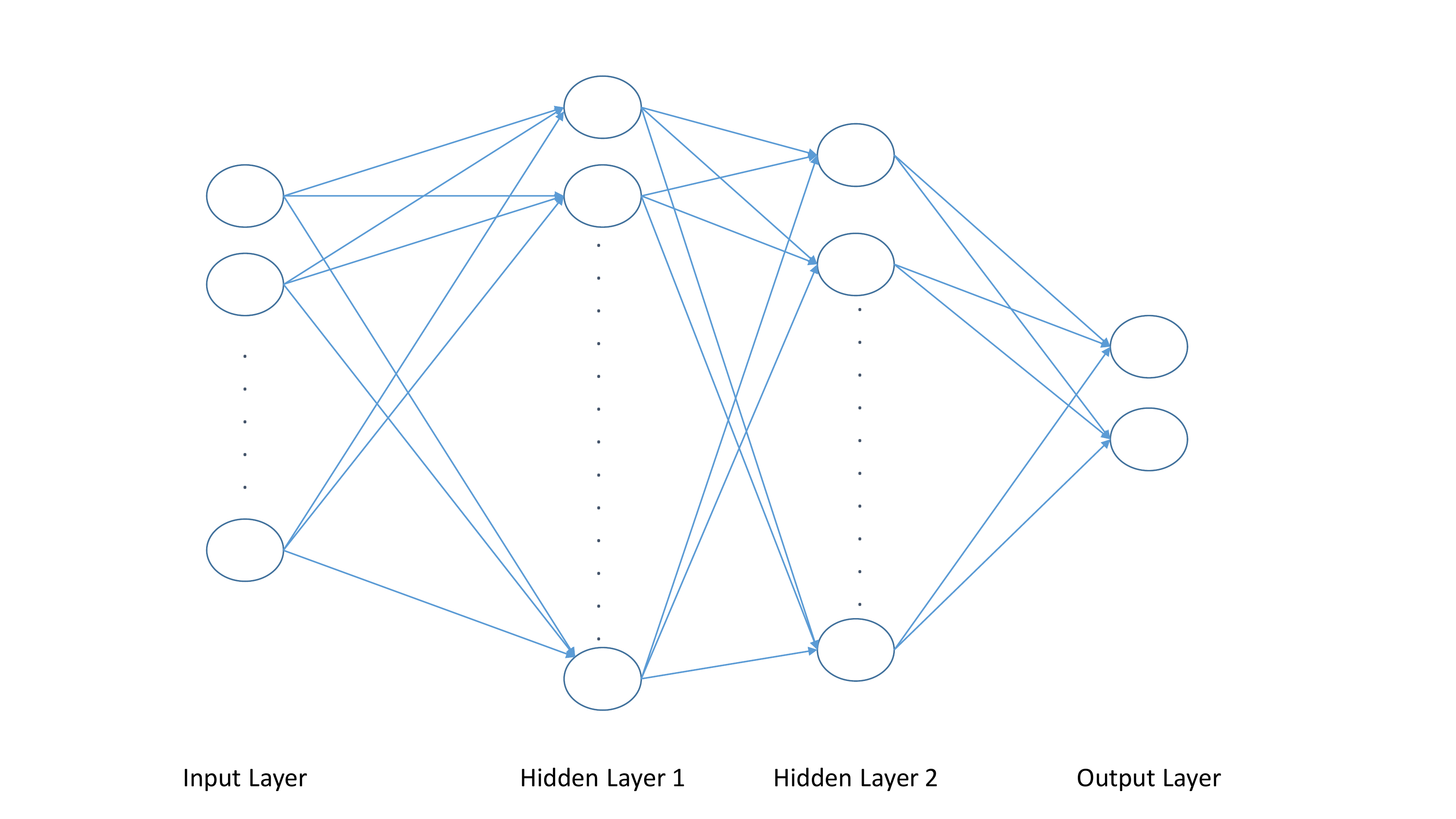}
\caption{The architecture of the MLP binary classifier. The network has two hidden layers and is fully connected.} \label{fig00}
\end{figure}

We train an MPL classifier that takes the weight vector as the input feature vector, and outputs a binary decision whether the displacement is purely axial or not. Figure \ref{fig00} shows the architecture of the used MLP model, which consists of an input layer, two hidden layers and an output layer. Our model is relatively simple due to having a low-dimensional input vector. The training is done by minimizing the mis-classification error using the cross-entropy loss function, and backpropagation is used to calculate the gradients. The applied optimization technique is the Adam optimizer \cite{ref_article9} with a learning rate of $1\mathrm{e}{-3}$. The MLP code is written in Python using Keras \cite{ref_url1}. 

\subsection{Data Collection}
\subsubsection{PCA Model}
For our training data, we collected
3,163 RF frames from 3 different CIRS phantoms (Norfolk, VA), namely Models
040GSE, 039 and 059 at different locations at Concordia University’s PERFORM Centre using a 12R Alpinion (Bothell, WA) ultrasound machine with an L3-12H high density linear array
probe. The center frequency is 8.5 MHz and the sampling 
frequency is 40 MHz. We allowed both in-plane and out-of-plane motion during collecting the data, where the probe could move in the 6 degrees of freedom (DOF). In addition, we have access to 420 RF frames collected from 4 patients undergoing liver ablation, where testing is done on only one of them.
The choice of the number of principal components was made so as to represent the displacement image in a simpler form while keeping most of the variance of the data. We chose $N=12$ which captures $95\%$ of the variance present in the original data using only a 12-dimentional feature vector.
\subsubsection{MLP Classifier}
We trained our model using 1,012 pairs of frames from the \textit{in-vivo} liver data through different combinations where each frame is paired with the nearest 16 frames forming 16 different pairs. We used 80\% of the data for training and 20\% for validation. Testing was done on a completely different dataset to ensure generalization. It is important to note that the ground truth (i.e. high or low quality strain image) was obtained by Abdelrahman Zayed through manual inspection of the strain image obtained using the Global Ultrasound Elastography technique~\cite{ref_article11}. The criteria for labelling the output as a good strain image were visual clarity and the ability to distinguish the inclusion from the surrounding tissue.

\section{Results}
We set $p=5$ RF lines as trials showed us that choosing a value for $p$ more than 5 would not improve the quality of the strain image \cite{ref_article12}. The number of hidden units in the MLP classifier is a hyperparameter that is chosen in a way so as to have the highest accuracy on the validation data.
The first and second hidden layers contain 64 and 32 hidden units respectively with a Rectified Linear Unit (ReLU) as the activation function. The output layer has two neurons with a softmax activation function.

For the PCA model, the unoptimized MATLAB code takes 5 hours to train the model, but it is only done once. During test time, extracting the features for two very large RF images of size $2304 \times 384$ using the procedure in Algorithm 1 takes $262$ ms on a 7th generation 3.4 GHz Intel core i7. As for the MLP classifier, training takes $5.57$ seconds after extracting the features from all the training dataset. For testing, our model takes only $1.9$ ms to choose the best frame by searching in a window composed of the nearest 16 frames (8 frames before and 8 frames after the desired frame), assuming that feature extraction is already done for the test dataset. 

Our model is tested on both tissue-mimicking phantom data and \textit{in-vivo} liver data. In order to be able to accurately measure the improvement in the quality of the strain image, we use two quality metrics which are the signal to
noise ratio (SNR) and contrast to noise ratio (CNR) \cite{ref_article13}, calculated as follows:
\begin{equation}
CNR=\frac{C}{N}=\sqrt{\frac{2(\bar{s_{b}}-\bar{s_{t}})^2}{\sigma_{b}^2 + \sigma_{t}^2} }, SNR=\frac{\bar{s}}{\sigma}
\end{equation}\\
where $ \bar{s_{t}}$ and $ \sigma_{t}^2$ are the strain average and variance of the target window (as shown in Figures \ref{fig1} and \ref{fig3}), $ \bar{s_{b}}$ and $ \sigma_{b}^2$ are the strain average and variance of the background window respectively. We use the background window for SNR calculation (i.e. $\bar{s}$=$ \bar{s_{b}}$ and $\sigma $=$ \sigma_{b}$). The background window is chosen in uniform areas. For the target window, we selected a window that lies completely inside the inclusion to show the contrast.

\subsection{Phantom Results} 
\begin{figure}[]
\begin{center}
\subfigure[B-mode]{\includegraphics[height=4 cm,width=5 cm]{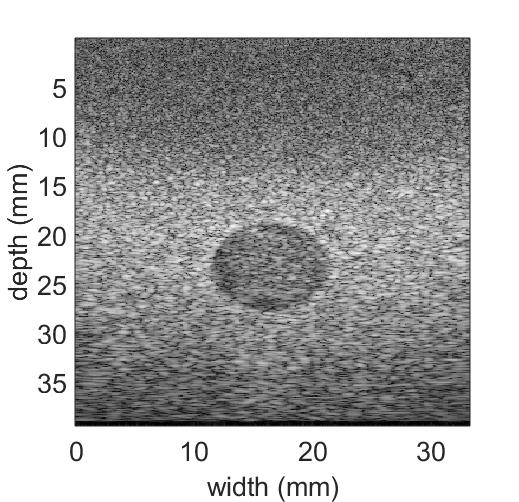}}
\subfigure[Strain from Skip 1 method]{\includegraphics[height=4 cm,width=5 cm]{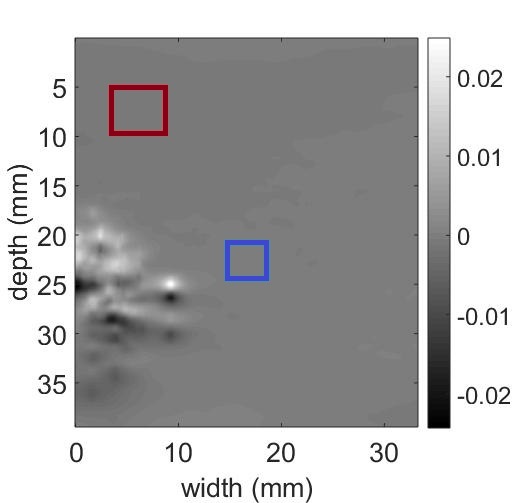}}\\
\subfigure[Strain from Skip 2 method]{\includegraphics[height=4 cm,width=5 cm]{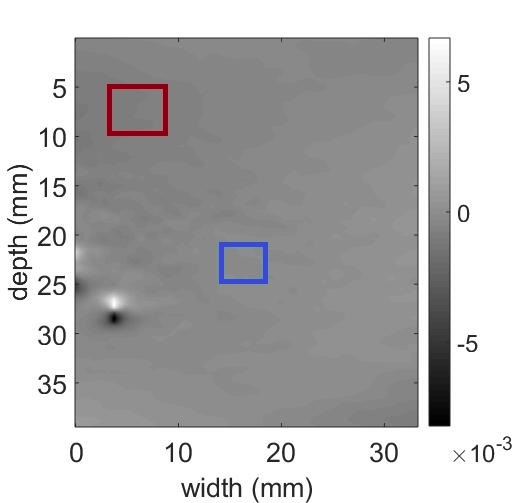}}
\subfigure[Strain from Skip 3 method]{\includegraphics[height=4 cm,width=5 cm]{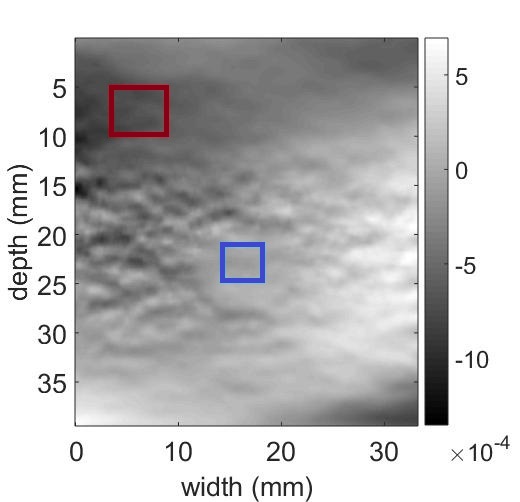}}
\subfigure[Strain from our method]{\includegraphics[height=4 cm,width=5 cm]{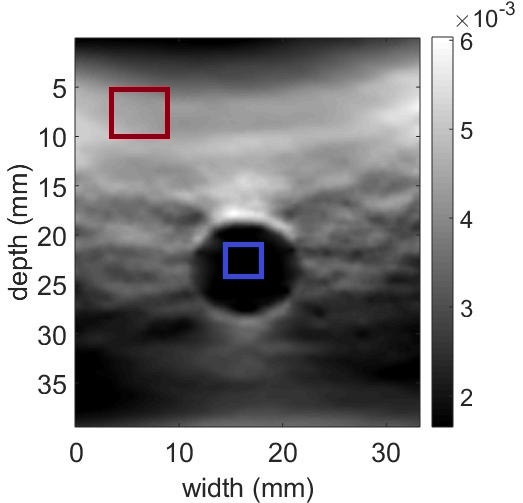}}
\end{center}%
\caption{The B-mode ultrasound and axial strain image for the phantom experiment.} \label{fig1}
\end{figure}

\begin{figure}[]
\begin{center}
\subfigure{\includegraphics[height=4 cm,width=5 cm]{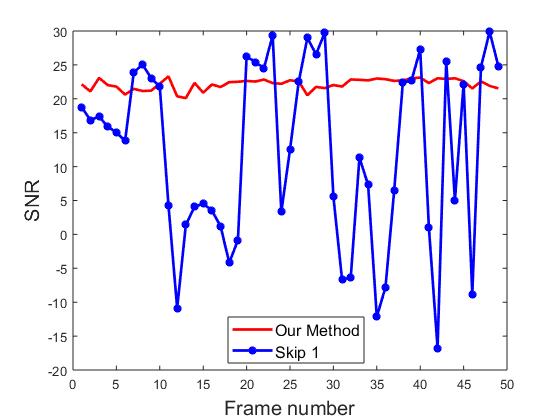}}
\subfigure{\includegraphics[height=4 cm,width=5 cm]{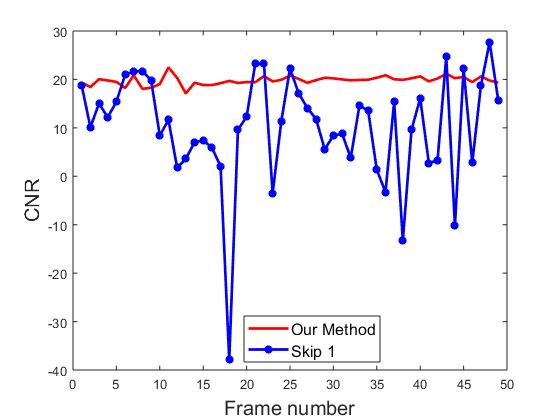}}\\
\subfigure{\includegraphics[height=4 cm,width=5 cm]{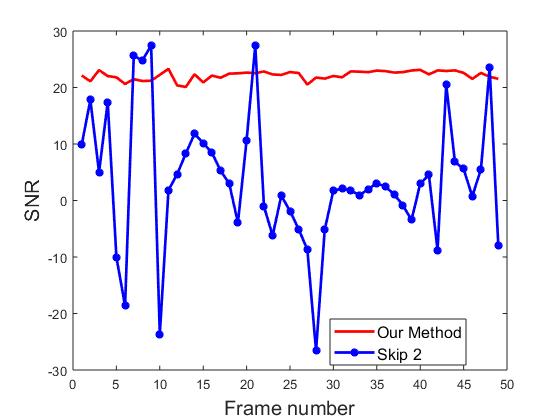}}
\subfigure{\includegraphics[height=4 cm,width=5 cm]{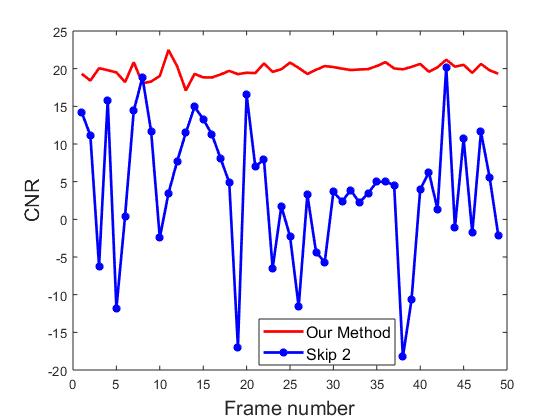}}\\
\subfigure{\includegraphics[height=4 cm,width=5 cm]{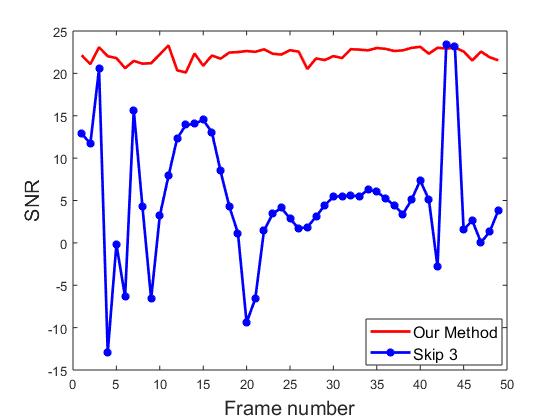}}
\subfigure{\includegraphics[height=4 cm,width=5 cm]{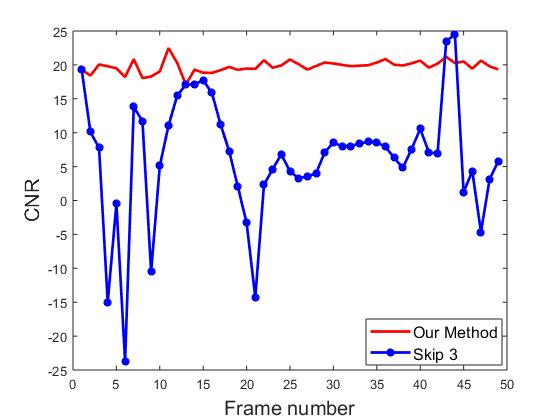}}
\end{center}%
\caption{A comparison between the SNR and CNR of the automatic frame selection and the fixed skip frame pairing for the phantom experiment. Rows 1 to 3 show the results for skipping 1 to 3 frames respectively.} \label{fig2}
\end{figure}

\begin{figure}[H]
\begin{center}
\subfigure[B-mode]{\includegraphics[height=4 cm,width=5 cm]{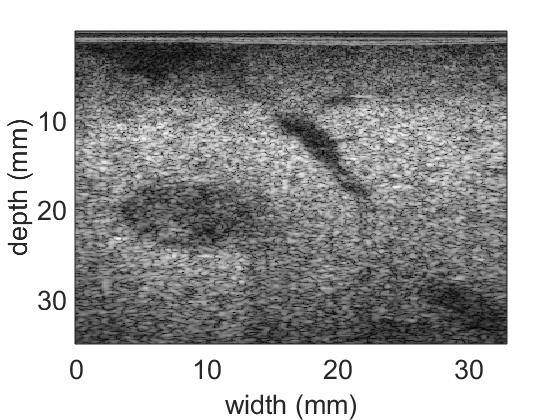}}
\subfigure[Strain from Skip 1 method]{\includegraphics[height=4 cm,width=5 cm]{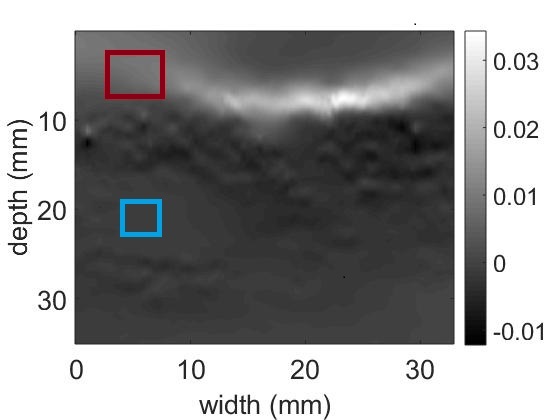}}\\
\subfigure[Strain from Skip 2 method]{\includegraphics[height=4 cm,width=5 cm]{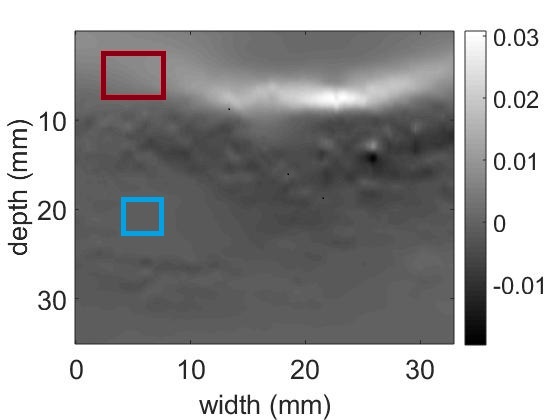}}
\subfigure[Strain from Skip 3 method]{\includegraphics[height=4 cm,width=5 cm]{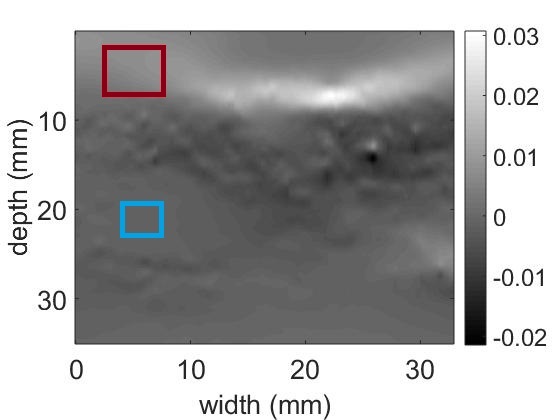}}
\subfigure[Strain from our method]{\includegraphics[height=4 cm,width=5 cm]{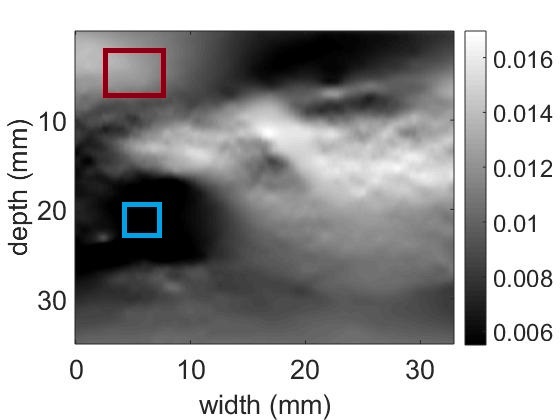}}
\end{center}%
\caption{The B-mode ultrasound and axial strain image for the \textit{in-vivo} experiment.} \label{fig3}
\end{figure}

We used data acquired from the CIRS elastography phantom Model 059 at a center frequency of 10 MHz and sampling
frequency of 40 MHz using the 12R Alpinion E-Cube ultrasound machine.
Fig.~\ref{fig1} shows the B-mode image as well as the axial strain images calculated using both our method and the fixed skip frame pairing.
Fig.~\ref{fig2} shows the SNR and CNR of the axial strain images calculated from the same experiment. It is clear that our automatic frame selection substantially outperforms simply skipping one, two or three frames. Table~\ref{tab1} summarizes the data in Fig.~\ref{fig2} by computing the average and standard deviation of the SNR and CNR.

\subsection{\textit{In-vivo} data}

Our \textit{in-vivo} results were obtained from one patient undergoing open surgical radiofrequency thermal ablation for
primary or secondary liver cancers. The data was acquired
at Johns Hopkins Hospital, with full details of the data collection protocol outlined in \cite{ref_article14}. Fig.~\ref{fig3} shows the B-mode image as well as the axial strain images using both our method and the fixed skip frame pairing.
Table~\ref{tab2} shows the average and standard deviation of the SNR and CNR of the axial strain images computed from the same experiment. As observed in the phantom experiment, automatic frames selection substantially improves the quality of the strain images.

\begin{table}[]
\centering
\caption{A comparison between SNR and CNR of the automatic frame selection and the fixed skip frame pairing for the phantom experiment. The numbers for each method show average $\pm$ standard deviation. }\label{tab1}
\begin{tabular}{|l|l|l|}
\hline
Method used & \,\,\,\,\,\,\,\,\,\,\,\,\,SNR & \,\,\,\,\,\,\,\,\,\,\,\,CNR\\
\hline
Skip 1 & {12.27\, $\pm   \,\,13.31$} &{ 10.11 $\pm \,\,11.36$}\\
Skip 2 & {\, 3.54 \,$\pm \,\,11.78$} &{\, \,3.80   \,$\pm \,\,\,\,\,8.92$}\\
Skip 3 & {\, 5.24 \,$\pm \,\,\,\,\,7.45$} &{\, \,6.34  \,$\pm \,\,\,\,\,9.09$} \\
Our method & {\textbf{22.15\,$\pm \,\,\,\textbf{0.79}$}} & \textbf{19.77$\,\,\pm \,\,\,\,\,\,\textbf{0.9}$}\\
\hline
\end{tabular}
\end{table}

\begin{table}[]
\centering
\caption{A comparison between the SNR and CNR of the automatic frame selection and the fixed skip frame pairing for the \textit{in-vivo} experiment.  The numbers for each method show average $\pm$ standard deviation.}\label{tab2}
\begin{tabular}{|l|l|l|}
\hline
Method used & \,\,\,\,\,\,\,\,\,\,\,\,SNR & \,\,\,\,\,\,\,\,\,\,\,\,CNR\\
\hline
Skip 1 & { 13.87 $\pm \,\,6.23$} &{  12.92 $\pm \,\,\,\,5.21$}\\
Skip 2 & { 13.60 $\pm \,\,7.11$} &{\, \,\,5.30 $\pm\,20.68$} \\
Skip 3 & { 13.54 $\pm \,\,8.74$} &{ 11.05 $\pm \,\,\,\,8.52$} \\
Our method & {\textbf{21.25 $\pm\,\textbf{2.23}$}} & \textbf{17.12 $\pm \,\,\textbf{3.22}$}\\
\hline
\end{tabular}
\end{table}

%
%

%
%
%
%

\section{Conclusion}
In this work, we presented a novel approach for real-time automatic selection of pairs of RF frames used to calculate the axial strain image. Our method is easy to use as it does not require any additional hardware. In addition, it is very computationally efficient and runs in less than 2 $ms$, and as such, can be used to test many pairs of RF frames in a short amount of time. Given that ultrasound frame rate is very high, and that there exist many combinations of two frames, this low computational complexity is of paramount practical importance. Our method can be used commerially where for each input RF frame, we choose the best possible frame to be paired with it among the collected frames.

\section*{Acknowledgements}
The authors would like to thank the principal investigators at Johns Hopkins Hospital Drs. E. Boctor, M. Choti and G. Hager for providing us with the \textit{in-vivo} liver data. 

\end{document}